\documentclass[11pt]{article}
\usepackage{Files/fa2025}
\usepackage{amsmath}
\usepackage{amssymb}
\usepackage{cite}
\usepackage{url}
\usepackage{graphicx}
\usepackage{color}
\usepackage{siunitx}
\usepackage[utf8]{inputenc}
\usepackage{enumitem}

\title{Optimal Pairwise Comparison Procedures for Subjective Evaluation}

\multauthor
{Jack Webb$^{1*}$ \hspace{1cm} Lorenzo Picinali $^1$} { 
  $^1$ Dyson School of Design Engineering, Imperial College London, United Kingdom\\
\correspondingauthor{j.webb24@imperial.ac.uk}{Webb \& Picinali.}
}

\begin{document}

\maketitle
\begin{abstract}
Audio signal processing algorithms are frequently assessed through subjective listening tests in which participants directly score degraded signals on a unidimensional numerical scale. However, this approach is susceptible to inconsistencies in scale calibration between assessors. Pairwise comparisons between degraded signals offer a more intuitive alternative, eliciting the relative scores of candidate signals with lower measurement error and reduced participant fatigue. Yet, due to the quadratic growth of the number of necessary comparisons, a complete set of pairwise comparisons becomes unfeasible for large datasets. This paper compares pairwise comparison procedures to identify the most efficient methods for approximating true quality scores with minimal comparisons. A novel sampling procedure is proposed and benchmarked against state-of-the-art methods on simulated datasets. Bayesian sampling produces the most robust score estimates among previously established methods, while the proposed procedure consistently converges fastest on the underlying ranking with comparable score accuracy.

\end{abstract}

\section{Introduction}\label{sec:introduction}
Subjective assessments of processed audio signals can be obtained through direct numerical ratings, or implicit ratings derived from pairwise comparisons. In the latter case, participants compare two audio signals and select the one they prefer or perceive as higher on a specific quality dimension, such as brightness. While direct ratings provide an immediate, fine-graded score, they also suffer from noisiness due to the difficulty of ensuring that rating scales are understood uniformly between participants \cite{li:18:scale}. Pairwise comparisons generally offer higher sensitivity and reduced noise \cite{shah:16}, as well as more efficient testing procedures due to the low complexity of the comparison task \cite{stewart:05}.

Several psychoacoustic studies have deployed pairwise comparisons \cite{burke:20, van:20, volandri:18, moore:11}, often via round-robin tournaments where each stimulus is compared against all others. However, a round-robin tournament between $n$ stimuli requires $n(n-1)/2$ comparisons, which quickly becomes impractical for large $n$. Furthermore, pairwise comparisons are still subject to noise from differences in observer expertise and preferences, random errors and experimental biases.

Multiple rounds of comparisons can mitigate these issues, but efficient sampling is essential to reduce the number of necessary comparisons while maintaining accuracy. This paper compares several canonical and state-of-the-art sampling methods, introducing a novel sorting approach, Sort-MST. Preliminary results indicate that Sort-MST yields the strongest ranking performance with score accuracies comparable to optimal Bayesian sampling approaches.

\hfill \break
\noindent Recent studies have focused on ranking the top-$\kappa$ stimuli from a larger set \cite{meyer:24}. In audio quality evaluation, it is often preferable to both rank all stimuli, and obtain accurate quality scores for each stimulus.

Extracting scores from pairwise comparison data requires a model to relate the observer responses to an arbitrary quality scale of interest. Two of the most common models are the Bradley-Terry (BT) model \cite{bradley:52}, and Thurstone's model \cite{thurstone:27}. Both models parameterise the $n$ stimuli as having underlying true scores $s = \{s_1,...,s_n\}, s \in \mathbb{R}$ on a given subjective scale such as quality, distortion or loudness. The BT model defines the probability of stimulus $i$ winning over stimulus $j$, $P_{ij}$, as: 

\begin{equation}
    P_{ij} = \frac{e^{s_i}}{e^{s_i} + e^{s_j}}
\end{equation}

\noindent Thurstone's model assumes each score $s_i$ is normally distributed with mean $\mu _i$ and standard deviation $\sigma_i$, giving $P_{ij}$:

\begin{equation}
    P_{ij} = \Phi\left(\frac{\mu_i - \mu_j}{\sqrt{\sigma_i^2 + \sigma_j^2}}\right)
\end{equation}

\noindent where $\Phi$ is the cumulative standard normal distribution function. Both models can be considered examples of generalised linear models, where the BT model is the logistic case, and Thurstone's model is the probit case. Estimated quality scores $\hat{s_i}$ can be calculated from comparison data using maximum likelihood estimation (MLE). This work focuses on the BT model, examining the robustness of different sampling algorithms against random judgment errors and natural variances in observer expertise that occur in a real pairwise comparison context. 

\section{Methods}

Six algorithms are implemented in MATLAB and investigated through computational simulation.

\subsection{Standard procedures}
Standard pairwise comparison procedures follow fixed rules for selecting comparisons. Here, we include:

\begin{enumerate}[wide, labelwidth=0pt, labelindent=0pt]
    \item \textit{Random sampling}. This acts as a baseline for minimum performance.
    \item \textit{Single-elimination knockout tournament}. A standard knockout (KO) tournament is implemented. Tournament trees are randomly initialised and run until the tournament is complete. New tournaments are initiated until the maximum number of comparisons is reached.
\end{enumerate}

\subsection{Active sampling}
Active sampling involves adaptively selecting comparisons based on previous comparison results. We examine sorting and information gain approaches.

\subsubsection{Sorting methods}
Following an initial set of comparisons, scores can be calculated by fitting BT or Thurstone models to the results. Sorting methods assume that pairs with the smallest difference in scores are the most informative to compare. Three algorithms are tested:

\begin{enumerate}[wide, labelwidth=0pt, labelindent=0pt, resume, start=3]
    
    \item \textit{Swiss tournament}. A Swiss-style tournament pairs stimuli with similar scores in each round without repeating pairings. This design has been shown to be the most efficient among standard tournament formats \cite{sziklai:22}. Our implementation uses opponent match win percentage (OMW) as a tiebreaker. To ensure robust rankings, tournaments run for $\lfloor\log_2(n)\rfloor + 2$ rounds before restarting.

    \item \textit{Tree selection}. The binary tree selection procedure detailed in \cite{meyer:24} is implemented as a modified knockout tournament. Though tree selection is primarily intended for top-$\kappa$ ranking, it is extended here to sort all stimuli.
    
    \item \textit{Sort-MST}. We propose a novel sorting method that leverages Elo scores, commonly used in chess, to pair stimuli with similar scores. Pairs are ranked by smallest difference in Elo scores, and the inverse of this ranking is used as edge weights to construct a minimum spanning tree. This returns a group of pairs in each round, ensuring a balanced design that avoids oversampling specific pairs.

\end{enumerate}

\subsubsection{Information gain methods}
Most optimal active sampling algorithms rely on information gain maximisation. The posterior distribution of quality scores is calculated at each stage, and the next pair is selected to maximise the information gained by observing the result of that comparison. Often, the information gain is calculated using the Kullback-Leibler (KL) divergence between the current distribution and the new distribution given the outcome of a potential future comparison, under a Bayesian statistical model. 

\begin{enumerate}[wide, labelwidth=0pt, labelindent=0pt, resume, start=6]
    \item \textit{Hybrid-MST}. Hybrid-MST and related approaches \cite{li:18, mikhailiuk:21} are fully Bayesian active sampling procedures that have recently demonstrated optimal performance. We implement Hybrid-MST to indicate the performance of state-of-the-art methods.
\end{enumerate}

\subsection{Procedure}

Monte Carlo simulations are performed on simulated datasets of 8, 16, and 32 stimuli. Each stimulus $i$ is assigned a ground truth score, $ s_i $, randomly sampled from a uniform distribution $ U(0,5) $. Standard deviations in each score, $ \sigma_i $, are sampled from three distinct ranges: $ U(0,0.4) $, $ U(0,0.7) $, and $ U(0,1.0) $. Smaller $\sigma$ values emulate clear distinctions between stimuli, while larger values simulate greater noise and observer uncertainty.

In a comparison between stimuli $i$ and $j$, observed scores $ r_i $ and $ r_j $ are drawn from normal distributions: $r_i \sim \mathcal{N}(s_i, \sigma_i), r_j \sim \mathcal{N}(s_j, \sigma_j)$. The match is won by the stimulus with the higher sampled score, unless a random judgment error flips the result. Judgment errors occur with probabilities $\epsilon \in \{0.1, 0.2, 0.3, 0.4\}$ to evaluate the robustness of each method to random noise. Each algorithm is tested over a range of 0 to 15 standard trials, where a standard trial consists of $n(n-1)/2$ comparisons. This process is repeated 100 times for each procedure.

\subsection{Evaluation}

We assess the algorithms by fitting BT scores every $n(n-1)/8$ comparisons. Ranking quality is assessed against the ground truth using Spearman's rank-order correlation coefficient (ROCC), while score accuracy is evaluated with Pearson's correlation coefficient (PCC) and root mean square error (RMSE). Since BT scores are scale-invariant, the estimated scores are aligned with the ground truth scores using sigmoid regression before RMSE is calculated, ensuring fair assessments.

\section{Results}\label{sec:results}

While the biggest factors in performance are the number of stimuli ($n$) and the noiseness of the comparisons ($\sigma$, $\epsilon$), there are clear differences between the various methods. Bayesian methods, such as Hybrid-MST, are optimised to achieve the lowest RMSE, and consequently demonstrate high correlations with ground truth scores across different $n$ and varying $\sigma$ and $\epsilon$. However, the proposed method, Sort-MST, obtains comparable PCC to Hybrid-MST in a range of conditions. With $n \geq 16$, the Swiss system also achieves strong correlations (Fig. 1).

\begin{figure}[ht]
 \includegraphics[width=8cm]{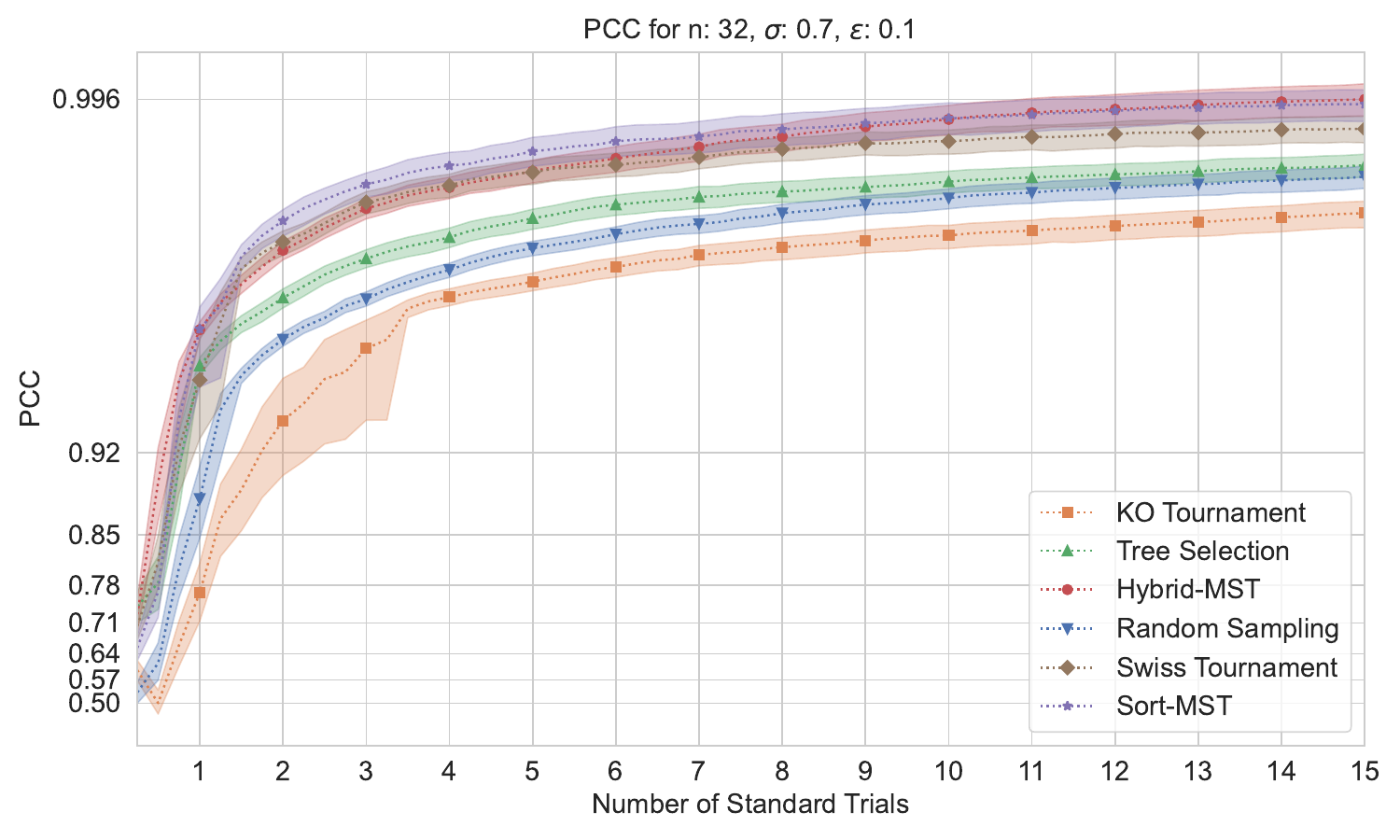}
 \includegraphics[width=8cm]{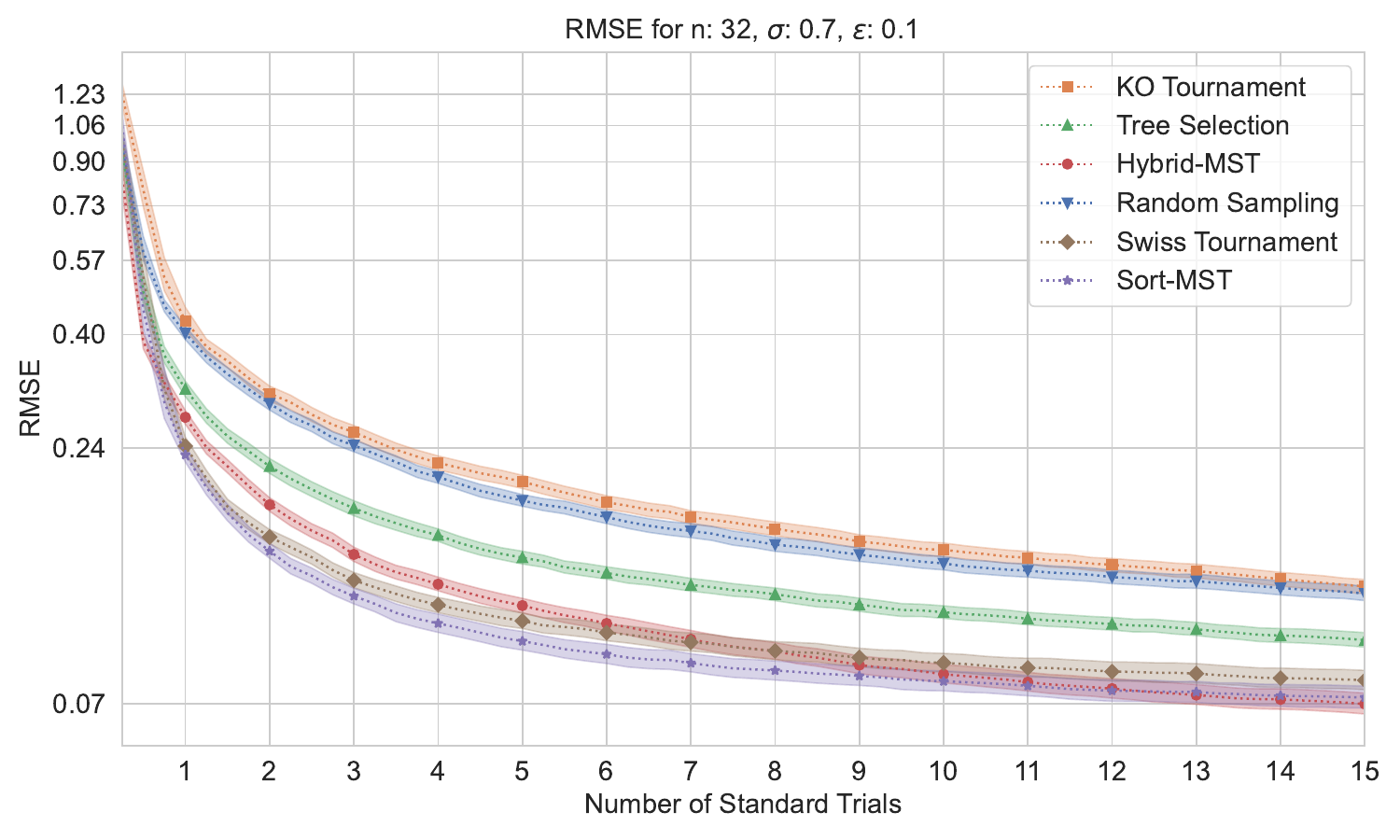}
 \caption{Mean PCC for and RMSE $n=32$. All graphs show 95\% confidence intervals bootstrapped over results from 100 repeats. Unless stated otherwise, all graphs show $\sigma \sim U(0,0.7)$, $\epsilon = 0.1$.}
 \label{fig:fig 1}
\end{figure}

Sort-MST also consistently achieves the highest ranking accuracy across different conditions (Fig. 2). Utilising a minimum spanning tree prevents imbalanced designs, meaning the algorithm can obtain accurate rankings without excessive distortions in PCC. By contrast, repeated KO tournaments yield outcomes equal to or worse than random sampling. Tree selection is generally an improvement on random sampling, but rarely performs optimally.

\begin{figure}[ht]
 \includegraphics[width=8cm]{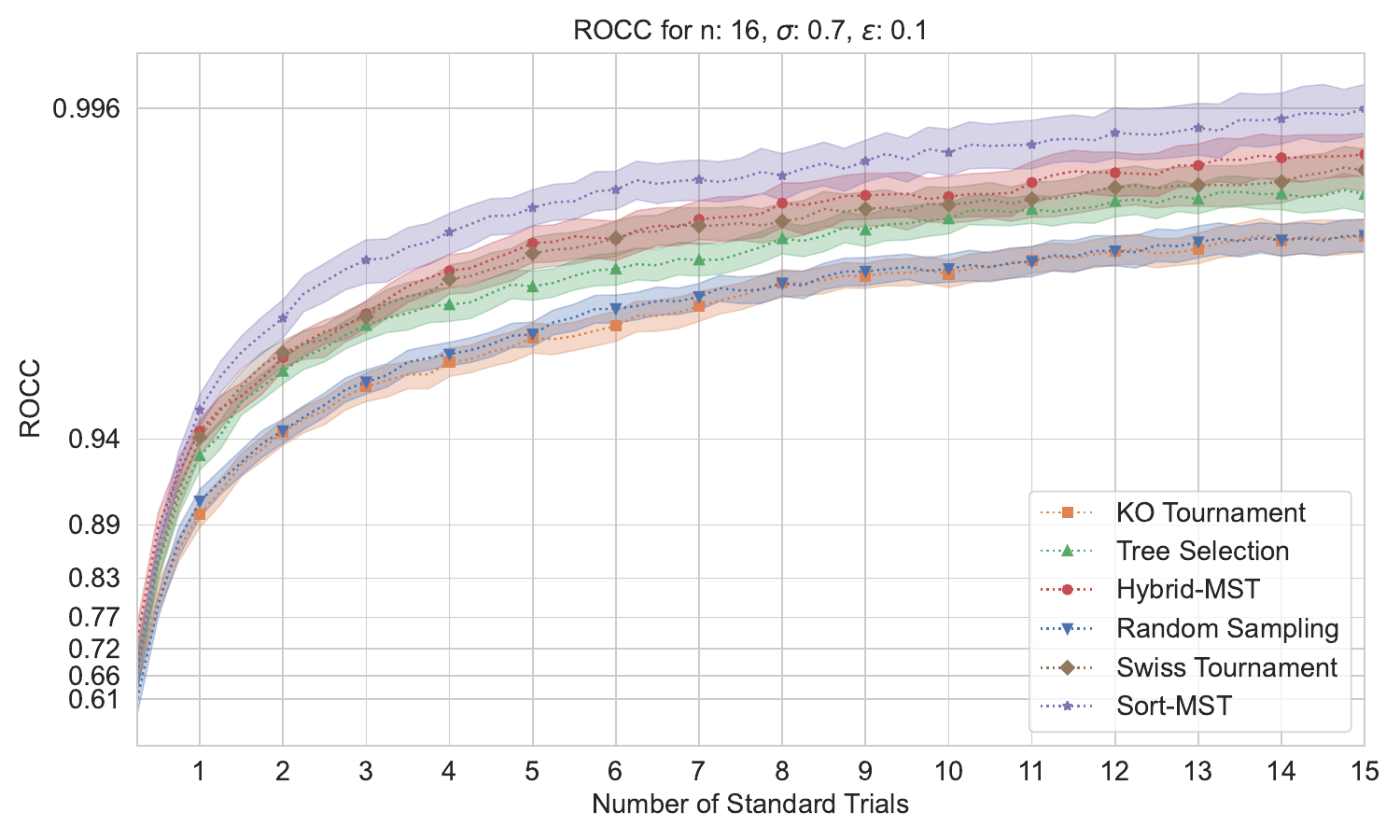}
 \caption{Mean ROCC for $n=16$.}
 \label{fig:fig 2}
\end{figure}

In conditions where uncertainty in relative stimulus strengths is high, sorting algorithms can outperform Bayesian methods (Fig. 3). Increased noisiness in pairwise comparison outcomes may lead to misleading information gain predictions. In this case, methods that do not explicitly model score uncertainties — such as the Swiss system or Sort-MST — may perform significantly better.

\begin{figure}[ht]
 \includegraphics[width=8cm]{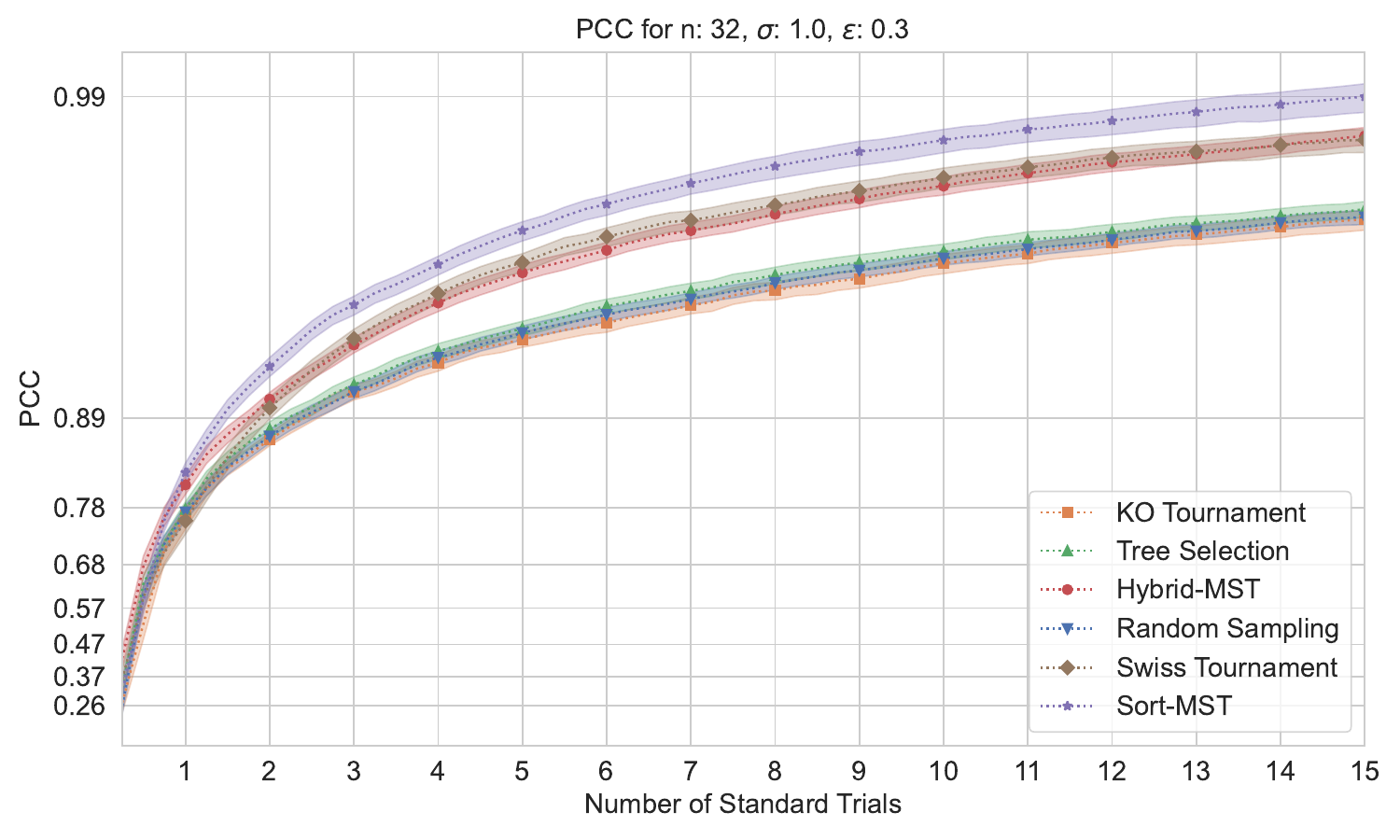}
 \caption{Mean PCC for $n=32$, $\sigma \sim U(0,1.0) $, $\epsilon = 0.3$}
 \label{fig:fig 3}
\end{figure}

\section{Conclusion}

Various pairwise comparison procedures are assessed using Monte Carlo simulations across a range of conditions. A novel sorting algorithm, Sort-MST, is shown to achieve the strongest rankings, with comparable score accuracy to Hybrid-MST, a Bayesian sampling approach that tends to yield the most robust scores. When uncertainty in the relative strengths of the stimuli is high, sorting methods can even outperform Bayesian sampling, with decreased computational expense. Future listening tests will validate these methods in real-world scenarios and investigate how scores can be aggregated across different observers.

\section{Acknowledgements}

This work was supported by an EPSRC CASE conversion PhD scholarship, co-funded by Sonova.

\bibliography{Optimal_Pairwise_Comparisons_FA2025}

\end{document}